\documentclass[aip,graphicx]{revtex4-1}

\draft % marks overfull lines with a black rule on the right

\begin{document}

\title{Euler observers in geometrodynamics} %Title of paper

\author{Alcides Garat}
%\thanks{}
\affiliation{1. Instituto de F\'{\i}sica, Facultad de Ciencias,
Igu\'a 4225, esq. Mataojo, Montevideo, Uruguay.}

\date{December 14th, 2011}

\begin{abstract}
Euler observers are a fundamental tool for the study of spacetime evolution. Cauchy surfaces are evolved through the use of hypersurface orthogonal fields and their relationship to coordinate observers, that enable the use of already developed algorithms. In geometrodynamics new tetrad vectors have been introduced with outstanding simplifying properties. We are going to use these already introduced tetrad vectors in the case where we consider a curved four dimensional Lorentzian spacetime with the presence of electromagnetic fields. These Einstein-Maxwell geometries will provide the new tetrad that we are going to use in order to develop an algorithm to produce Cauchy evolution with additional simplifying properties.
\end{abstract}


\maketitle %\maketitle must follow title, authors, abstract and \pacs
\section{Introduction}
\label{introduction}

The new tetrads introduced in \cite{A}, yield maximum simplification in the expression of a non-null electromagnetic field in a curved spacetime. We present first, the four tetrad vectors introduced in paper \cite{A} that locally and covariantly diagonalize the electromagnetic stress-energy tensor and define at every point in spacetime the blades one and two.

\begin{eqnarray}
V_{(1)}^{\alpha} &=& \xi^{\alpha\lambda}\:\xi_{\rho\lambda}\:X^{\rho}
\label{V1}\\
V_{(2)}^{\alpha} &=& \sqrt{-Q/2} \:\: \xi^{\alpha\lambda} \: X_{\lambda}
\label{V2}\\
V_{(3)}^{\alpha} &=& \sqrt{-Q/2} \:\: \ast \xi^{\alpha\lambda} \: Y_{\lambda}
\label{V3}\\
V_{(4)}^{\alpha} &=& \ast \xi^{\alpha\lambda}\: \ast \xi_{\rho\lambda}
\:Y^{\rho}\ ,\label{V4}
\end{eqnarray}

where $Q=\xi_{\mu\nu}\:\xi^{\mu\nu}=-\sqrt{T_{\mu\nu}T^{\mu\nu}}$ according to equations (39) in \cite{MW}. $Q$ is assumed not to be zero, because we are dealing with non-null electromagnetic fields. The first two (\ref{V1}-\ref{V2}) eigenvectors of the stress-energy tensor with eigenvalue $Q/2$, the last two (\ref{V3}-\ref{V4}) with eigenvalue $-Q/2$. We briefly remind ourselves that the original expression for the electromagnetic stress-energy tensor $T_{\mu\nu}= f_{\mu\lambda}\:\:f_{\nu}^{\:\:\:\lambda}
+ \ast f_{\mu\lambda}\:\ast f_{\nu}^{\:\:\:\lambda}$ was given in terms of the electromagnetic tensor $f_{\mu\nu}$ and its dual $\ast f_{\mu\nu}={1 \over 2}\:\epsilon_{\mu\nu\sigma\tau}\:f^{\sigma\tau}$. After a local duality transformation,

\begin{equation}
f_{\mu\nu} = \xi_{\mu\nu} \: \cos\alpha +
\ast\xi_{\mu\nu} \: \sin\alpha\ ,\label{dr}
\end{equation}

where the local scalar $\alpha$ is the complexion, we are able to write the stress-energy in terms of the extremal field $\xi_{\mu\nu}$ and its dual. We can express the extremal field as,

\begin{equation}
\xi_{\mu\nu} = e^{-\ast \alpha} f_{\mu\nu}\ = \cos\alpha\:f_{\mu\nu} - \sin\alpha\:\ast f_{\mu\nu}.\label{dref}
\end{equation}

Extremal fields are essentially electric fields and they satisfy,

\begin{equation}
\xi_{\mu\nu} \ast \xi^{\mu\nu}= 0\ . \label{i0}
\end{equation}

Equation (\ref{i0}) is a condition imposed on (\ref{dref}) and then the explicit expression for the complexion emerges, $\tan(2\alpha) = - f_{\mu\nu}\:\ast f^{\mu\nu} / f_{\lambda\rho}\:f^{\lambda\rho}$. As antisymmetric fields in a four dimensional Lorentzian spacetime, the extremal fields also verify the identity,

\begin{eqnarray}
\xi_{\mu\alpha}\:\xi^{\nu\alpha} -
\ast \xi_{\mu\alpha}\: \ast \xi^{\nu\alpha} &=& \frac{1}{2}
\: \delta_{\mu}^{\:\:\:\nu}\ Q \ .\label{i1}
\end{eqnarray}

It can be proved that condition (\ref{i0}) and through the use of the general identity,

\begin{eqnarray}
A_{\mu\alpha}\:B^{\nu\alpha} -
\ast B_{\mu\alpha}\: \ast A^{\nu\alpha} &=& \frac{1}{2}
\: \delta_{\mu}^{\:\:\:\nu}\: A_{\alpha\beta}\:B^{\alpha\beta}  \ ,\label{ig}
\end{eqnarray}

which is valid for every pair of antisymmetric tensors in a four-dimensional Lorentzian spacetime \cite{MW}, when applied to the case $A_{\mu\alpha} = \xi_{\mu\alpha}$ and $B^{\nu\alpha} = \ast \xi^{\nu\alpha}$ yields the equivalent condition,

\begin{eqnarray}
\xi_{\alpha\mu}\:\ast \xi^{\mu\nu} &=& 0\ ,\label{i2}
\end{eqnarray}

which is equation (64) in \cite{MW}. It is evident that identity (\ref{i1}) is a special case of (\ref{ig}). The duality rotation given by equation (\ref{dr}) allows us to express the stress-energy tensor in terms of the extremal field,

\begin{equation}
T_{\mu\nu}=\xi_{\mu\lambda}\:\:\xi_{\nu}^{\:\:\:\lambda}
+ \ast \xi_{\mu\lambda}\:\ast \xi_{\nu}^{\:\:\:\lambda}\ .\label{TEMDR}
\end{equation}

With all these elements it becomes trivial to prove that the tetrad (\ref{V1}-\ref{V4}) is orthogonal and diagonalizes the stress-energy tensor (\ref{TEMDR}). We notice then that we still have to define the vectors $X^{\mu}$ and $Y^{\mu}$. Let us introduce some names. The tetrad vectors have two essential components. For instance in vector $V_{(1)}^{\alpha}$ there are two main structures. First, the skeleton, in this case $\xi^{\alpha\lambda}\:\xi_{\rho\lambda}$, and second, the gauge vector $X^{\rho}$. The gauge vectors it was proved in manuscript \cite{A} could be anything that does not make the tetrad vectors trivial. That is, the tetrad (\ref{V1}-\ref{V4}) diagonalizes the stress-energy tensor for any non-trivial gauge vectors $X^{\mu}$ and $Y^{\mu}$. It was therefore proved that we can make different choices for $X^{\mu}$ and $Y^{\mu}$. In geometrodynamics, the Maxwell equations,

\begin{eqnarray}
f^{\mu\nu}_{\:\:\:\:\:;\nu} &=& 0 \label{L1}\nonumber\\
\ast f^{\mu\nu}_{\:\:\:\:\:;\nu} &=& 0 \ , \label{L2}
\end{eqnarray}

are telling us that two potential vector fields $A_{\nu}$ and $\ast A_{\nu}$ exist,

\begin{eqnarray}
f_{\mu\nu} &=& A_{\nu ;\mu} - A_{\mu ;\nu}\label{ER}\nonumber\\
\ast f_{\mu\nu} &=& \ast A_{\nu ;\mu} - \ast A_{\mu ;\nu} \ .\label{DER}
\end{eqnarray}

The symbol $``;''$ stands for covariant derivative with respect to the metric tensor $g_{\mu\nu}$. We can define then, a tetrad,

\begin{eqnarray}
U^{\alpha} &=& \xi^{\alpha\lambda}\:\xi_{\rho\lambda}\:A^{\rho} \:
/ \: (\: \sqrt{-Q/2} \: \sqrt{A_{\mu} \ \xi^{\mu\sigma} \
\xi_{\nu\sigma} \ A^{\nu}}\:) \label{U}\\
V^{\alpha} &=& \xi^{\alpha\lambda}\:A_{\lambda} \:
/ \: (\:\sqrt{A_{\mu} \ \xi^{\mu\sigma} \
\xi_{\nu\sigma} \ A^{\nu}}\:) \label{V}\\
Z^{\alpha} &=& \ast \xi^{\alpha\lambda} \: \ast A_{\lambda} \:
/ \: (\:\sqrt{\ast A_{\mu}  \ast \xi^{\mu\sigma}
\ast \xi_{\nu\sigma}  \ast A^{\nu}}\:)
\label{Z}\\
W^{\alpha} &=& \ast \xi^{\alpha\lambda}\: \ast \xi_{\rho\lambda}
\:\ast A^{\rho} \: / \: (\:\sqrt{-Q/2} \: \sqrt{\ast A_{\mu}
\ast \xi^{\mu\sigma} \ast \xi_{\nu\sigma} \ast A^{\nu}}\:) \ .
\label{W}
\end{eqnarray}

The four vectors (\ref{U}-\ref{W}) have the following algebraic properties,

\begin{equation}
-U^{\alpha}\:U_{\alpha}=V^{\alpha}\:V_{\alpha}
=Z^{\alpha}\:Z_{\alpha}=W^{\alpha}\:W_{\alpha}=1 \ .\label{TSPAUX}
\end{equation}

Using the equations (\ref{i1}-\ref{i2}) it is simple to prove that (\ref{U}-\ref{W}) are orthonormal. When we make the transformation,

\begin{eqnarray}
A_{\alpha} \rightarrow A_{\alpha} + \Lambda_{,\alpha}\ , \label{G1}
\end{eqnarray}

$f_{\mu\nu}$ remains invariant, and the transformation,

\begin{eqnarray}
\ast A_{\alpha} \rightarrow \ast A_{\alpha} +
\ast \Lambda_{,\alpha}\ , \label{G2}
\end{eqnarray}

leaves $\ast f_{\mu\nu}$ invariant,
as long as the functions $\Lambda$ and $\ast \Lambda$ are
scalars. Schouten \cite{SCH} defined what he called, a two-bladed structure
in a spacetime \cite{SCH}. These blades are the planes determined by the pairs
($U^{\alpha}, V^{\alpha}$) and ($Z^{\alpha}, W^{\alpha}$).
It was proved in \cite{A} that the transformation (\ref{G1}) generates a ``rotation'' of the tetrad vectors ($U^{\alpha}, V^{\alpha}$) into ($\tilde{U}^{\alpha}, \tilde{V}^{\alpha}$) such that these ``rotated'' vectors ($\tilde{U}^{\alpha}, \tilde{V}^{\alpha}$) remain in the plane or blade one generated by ($U^{\alpha}, V^{\alpha}$). It was also proved in \cite{A} that the transformation (\ref{G2}) generates a ``rotation'' of the tetrad vectors ($Z^{\alpha}, W^{\alpha}$) into ($\tilde{Z}^{\alpha}, \tilde{W}^{\alpha}$) such that these ``rotated'' vectors ($\tilde{Z}^{\alpha}, \tilde{W}^{\alpha}$) remain in the plane or blade two generated by ($Z^{\alpha}, W^{\alpha}$).  For example, a boost of the two vectors $(U^{\alpha},\:V^{\alpha})$ on blade one, given in (\ref{U}-\ref{V}), by the ``angle'' $\phi$ can be written,

\begin{eqnarray}
U^{\alpha}_{(\phi)}  &=& \cosh(\phi)\: U^{\alpha} +  \sinh(\phi)\: V^{\alpha} \label{UT} \\
V^{\alpha}_{(\phi)} &=& \sinh(\phi)\: U^{\alpha} +  \cosh(\phi)\: V^{\alpha} \label{VT} \ .
\end{eqnarray}

There are also discrete transformations of vectors $(U^{\alpha},\:V^{\alpha})$ on blade one \cite{A}. The rotation of the two tetrad vectors $(Z^{\alpha},\:W^{\alpha})$ on blade two, given in (\ref{Z}-\ref{W}), by the ``angle'' $\varphi$, can be expressed as,

\begin{eqnarray}
Z^{\alpha}_{(\varphi)}  &=& \cos(\varphi)\: Z^{\alpha} -  \sin(\varphi)\: W^{\alpha} \label{ZT} \\
W^{\alpha}_{(\varphi)}  &=& \sin(\varphi)\: Z^{\alpha} +  \cos(\varphi)\: W^{\alpha} \label{WT} \ .
\end{eqnarray}

It is a simple exercise in algebra to see that the equalities $U^{[\alpha}_{(\phi)}\:V^{\beta]}_{(\phi)} = U^{[\alpha}\:V^{\beta]}$ and $Z^{[\alpha}_{(\varphi)}\:W^{\beta]}_{(\varphi)} = Z^{[\alpha}\:W^{\beta]}$ are true. These equalities are telling us that these antisymmetric tetrad objects are gauge invariant. We remind ourselves that it was proved in manuscript \cite{A} that the group of local electromagnetic gauge transformations is isomorphic to the local group LB1 of boosts plus discrete transformations on blade one, and independently to LB2, the local group of rotations on blade two. Equations (\ref{UT}-\ref{VT}) represent a local electromagnetic gauge transformation of the vectors $(U^{\alpha}, V^{\alpha})$. Equations (\ref{ZT}-\ref{WT}) represent a local electromagnetic gauge transformation of the vectors $(Z^{\alpha}, W^{\alpha})$. Written in terms of these tetrad vectors, the electromagnetic field is,

\begin{equation}
f_{\alpha\beta} = -2\:\sqrt{-Q/2}\:\:\cos\alpha\:\:U_{[\alpha}\:V_{\beta]} +
2\:\sqrt{-Q/2}\:\:\sin\alpha\:\:Z_{[\alpha}\:W_{\beta]}\ .\label{EMF}
\end{equation}

Having introduced the new tetrad we proceed in section \ref{surface} to introduce the hypersurface orthogonality condition and the algorithm to build the Eulerian vector fields. We use a metric with sign conventions $-+++$. If $F_{\mu\nu}$ is the electromagnetic field then $f_{\mu\nu}= (G^{1/2} / c^2) \: F_{\mu\nu}$ is the geometrized electromagnetic field.

%%%%%%%%%%%%%%%%%%%%%%%%%%%%%%%%%%%%%%%%%%%%%%%%%%%%%%%%%%%%%%%%%%%%%%%%%

\section{Euler vector fields}
\label{surface}

First we write the explicit equations satisfied by the hypersurface orthogonal\cite{MC}$^{,}$\cite{RW}$^{,}$\cite{SYO} unit vector fields $n_{\mu}\:n^{\mu} = -1$,

\begin{eqnarray}
n_{\alpha}\:n_{\beta;\gamma} + n_{\beta}\:n_{\gamma;\alpha} +  n_{\gamma}\:n_{\alpha;\beta}
- n_{\alpha}\:n_{\gamma;\beta} - n_{\gamma}\:n_{\beta;\alpha} -  n_{\beta}\:n_{\alpha;\gamma} = 0 \ .\label{hyper}
\end{eqnarray}

The Euler unit timelike vector field that we are going to find through our algorithm satisfying equation (\ref{hyper}) we are going to name $\hat{U}^{\mu}$. The other three  vectors in the new orthonormal tetrad are $\hat{V}^{\mu}$, $\hat{Z}^{\mu}$ and $\hat{W}^{\mu}$. Therefore the hypersurface orthogonal vector $\hat{U}^{\mu}$ is going to satisfy the equation,

\begin{eqnarray}
\hat{U}_{\alpha}\:\hat{U}_{\beta;\gamma} + \hat{U}_{\beta}\:\hat{U}_{\gamma;\alpha} +  \hat{U}_{\gamma}\:\hat{U}_{\alpha;\beta}
- \hat{U}_{\alpha}\:\hat{U}_{\gamma;\beta} - \hat{U}_{\gamma}\:\hat{U}_{\beta;\alpha} -  \hat{U}_{\beta}\:\hat{U}_{\alpha;\gamma} = 0 \ .\label{hyperhat}
\end{eqnarray}

If we project equation (\ref{hyperhat}) using the four tetrad vectors  ($\hat{U}^{\alpha}, \hat{V}^{\alpha}, \hat{Z}^{\alpha}, \hat{W}^{\alpha}$) we get only three meaningful equations,

\begin{eqnarray}
\hat{U}_{[\alpha ; \beta]} \: \hat{V}^{\alpha}\:Z^{\beta} &=& 0 \label{VZ}\\
\hat{U}_{[\alpha ; \beta]} \: \hat{V}^{\alpha}\:W^{\beta} &=& 0  \label{VW}\\
\hat{U}_{[\alpha ; \beta]} \: \hat{Z}^{\alpha}\:W^{\beta} &=& 0   \ . \label{ZW}
\end{eqnarray}

Equations (\ref{VZ}-\ref{ZW}) are three conditions on the vector field $\hat{U}^{\alpha}$. If we first perform a local boost like in equations (\ref{UT}-\ref{VT}) and then a rotation like in equations (\ref{ZT}-\ref{WT}) we would have two local scalars ($\phi, \varphi$) that can represent two of the three variables necessary to find a meaningful solution to equations (\ref{VZ}-\ref{ZW}). Next we perform a boost in the plane spanned by ($U^{\alpha}_{(\phi)}, W^{\alpha}_{(\varphi)}$),

\begin{eqnarray}
\hat{U}^{\alpha}  &=& \cosh(\psi)\:U^{\alpha}_{(\phi)}  +  \sinh(\psi)\:W^{\alpha}_{(\varphi)}  \label{UWf} \\
\hat{W}^{\alpha}  &=& \sinh(\psi)\:U^{\alpha}_{(\phi)}  +  \cosh(\psi)\:W^{\alpha}_{(\varphi)}  \label{UWs} \ .
\end{eqnarray}

In this way we introduce a third local scalar $\psi$ that completes the necessary three local variables that are going to be the solution to the system (\ref{VZ}-\ref{ZW}). The final orthonormal tetrad that has as a timelike vector field the hypersurface orthogonal vector field that we want as an input for our evolution algorithms is given by,

\begin{eqnarray}
\hat{U}^{\alpha}  &=& \cosh(\psi)\:U^{\alpha}_{(\phi)}  +  \sinh(\psi)\:W^{\alpha}_{(\varphi)}  \label{SFU} \\
\hat{V}^{\alpha} &=& V^{\alpha}_{(\phi)} \label{SFV} \\
\hat{Z}^{\alpha}  &=& Z^{\alpha}_{(\varphi)}\label{SFZ} \\
\hat{W}^{\alpha}  &=& \sinh(\psi)\:U^{\alpha}_{(\phi)}  +  \cosh(\psi)\:W^{\alpha}_{(\varphi)}  \label{SFW} \ .
\end{eqnarray}

We notice that in order for the algorithm to be meaningful the tetrad vector that has to involve after the three Lorentz transformations the three local scalars $(\phi, \varphi, \psi)$, is $\hat{U}^{\alpha}$. If, for instance, we introduce Lorentz transformations only involving the original vectors $(V^{\alpha}, Z^{\alpha}, W^{\alpha})$ then we would only obtain combinations of the original equations (\ref{VZ}-\ref{ZW}) and since these can be algebraically decoupled, we would not be introducing any new information. The new information comes from the inclusion of the three local scalars $(\phi, \varphi, \psi)$ inside the derivatives and this happens only through the vector $\hat{U}^{\alpha}$. As a last issue, we would like to study the contraction of the the tetrad vectors $(\hat{U}^{\alpha}, \hat{V}^{\alpha}, \hat{Z}^{\alpha}, \hat{W}^{\alpha})$ with the stress-energy tensor (\ref{TEMDR}).

\begin{eqnarray}
\hat{U}^{\alpha}\:T_{\alpha}^{\:\:\:\beta} &=& \frac{Q}{2}\:(\cosh(\psi)\:U^{\alpha}_{(\phi)} - \sinh(\psi)\:W^{\alpha}_{(\varphi)}) \label{EV1}\\
\hat{V}^{\alpha}\:T_{\alpha}^{\:\:\:\beta} &=& \frac{Q}{2}\:\hat{V}^{\alpha} \label{EV2}\\
\hat{Z}^{\alpha}\:T_{\alpha}^{\:\:\:\beta} &=& -\frac{Q}{2}\:\hat{Z}^{\alpha} \label{EV3}\\
\hat{W}^{\alpha}\:T_{\alpha}^{\:\:\:\beta} &=& -\frac{Q}{2}\:(- \sinh(\psi)\:U^{\alpha}_{(\phi)} + \cosh(\psi)\:W^{\alpha}_{(\varphi)})\ .\label{EV4}
\end{eqnarray}

Therefore, the only non-zero components of the stress-energy tensor when expressed in terms of the new tetrad are,

\begin{eqnarray}
\hat{U}^{\alpha}\:T_{\alpha}^{\:\:\:\beta}\:\hat{U}_{\beta} &=& \frac{Q}{2}\:(- \cosh^{2}(\psi) - \sinh^{2}(\psi)) \label{SE00}\\
\hat{V}^{\alpha}\:T_{\alpha}^{\:\:\:\beta}\:\hat{V}_{\beta} &=& \frac{Q}{2} \label{SE11}\\
\hat{Z}^{\alpha}\:T_{\alpha}^{\:\:\:\beta}\:\hat{Z}_{\beta} &=& -\frac{Q}{2} \label{SE22}\\
\hat{W}^{\alpha}\:T_{\alpha}^{\:\:\:\beta}\:\hat{W}_{\beta} &=& -\frac{Q}{2}\:(\sinh^{2}(\psi) +  \cosh^{2}(\psi)) \label{SE33}\\
\hat{U}^{\alpha}\:T_{\alpha}^{\:\:\:\beta}\:\hat{W}_{\beta} &=& \frac{Q}{2}\:(-\sinh(\psi)\:\cosh(\psi) - \cosh(\psi)\:\sinh(\psi))\ .\label{SE03}
\end{eqnarray}

We finally get,

\begin{eqnarray}
\hat{U}^{\alpha}\:T_{\alpha}^{\:\:\:\beta}\:\hat{U}_{\beta} &=& -\frac{Q}{2}\:\cosh(2\psi) \label{SE00F}\\
\hat{V}^{\alpha}\:T_{\alpha}^{\:\:\:\beta}\:\hat{V}_{\beta} &=& \frac{Q}{2} \label{SE11F}\\
\hat{Z}^{\alpha}\:T_{\alpha}^{\:\:\:\beta}\:\hat{Z}_{\beta} &=& -\frac{Q}{2} \label{SE22F}\\
\hat{W}^{\alpha}\:T_{\alpha}^{\:\:\:\beta}\:\hat{W}_{\beta} &=& -\frac{Q}{2}\:\cosh(2\psi) \label{SE33F}\\
\hat{U}^{\alpha}\:T_{\alpha}^{\:\:\:\beta}\:\hat{W}_{\beta} &=& -\frac{Q}{2}\:\sinh(2\psi)\ .\label{SE03F}
\end{eqnarray}

We can assert that in terms of the new tetrad, the stress-energy tensor acquires a very simple set of components with only one off-diagonal non-zero component. In this way the new tetrad provides both a hypersurface orthogonal congruence and a maximum simplification of the stress-energy tensor given that the tetrad that diagonalized the tensor underwent three Lorentz transformations. It is evident from expressions (\ref{SE00F}-\ref{SE03F}) that we recover the results for the old tetrad that diagonalizes the stress-energy tensor when we take the limit $\psi \rightarrow 0$.

\section{Conclusions}
\label{conclusions}

We are considering dynamical situations where gravitational and electromagnetic fields are evolving. The symmetries in the gauge theory of electromagnetic fields are understood through the isomorphisms proved in manuscript \cite{A} as local Lorentz transformations on either blade one or two. Local groups that we named LB1 and LB2. New local tetrad vectors transform inside these blades under the action of these groups. When an external agent to the preexisting geometry perturbes the original system, the local planes of symmetry are tilted with respect to the original ones. The symmetries are going to correspond to new local planes. The vectors that locally diagonalize the old stress-energy tensor will no longer diagonalize the new perturbed stress-energy tensor. We can specify the old and new tetrad vectors by two features. On one hand what we might call the tetrad vectors skeleton and on the other hand the gauge vectors. As an example of skeleton we can see for instance the $\xi^{\alpha\lambda}\:\xi_{\rho\lambda}$ in the vector $V_{(1)}^{\alpha}$. In the same vector the gauge vector would be $X^{\rho}$. Nonetheless, and this is an outstanding property of these new tetrads, the local tetrad structure in terms of skeletons, on one hand and gauge fields on the other will remain structure invariant after the ensuing perturbation. Even though the tetrad that diagonalizes the original stress-energy tensor is not the same as the new one that diagonalizes the perturbed stress-energy tensor, the tetrad vectors in both cases are locally, structure invariant. This simply occurs because we can always apply the duality transformation technique to the perturbed fields and obtain extremal fields for the perturbed electromagnetic field and so on. The new perturbed extremal fields allow for the construction of a new tetrad that diagonalizes the perturbed stress-energy tensor. There is a symmetry evolution, and we evaluate this evolution through the local plane symmetry evolution, or the evolution of blades one and two. In other words the local evolution of the groups LB1 and LB2. In our algorithm we carry out one boost on blade one, one rotation on blade two, and finally, one boost on the plane determined by the timelike vector on blade one and one of the two already rotated spacelike vectors on blade two. We introduce in this way the three local scalars necessary to solve the hypersurface orthogonality evolution problem. It is evident that with these new Euler observers we can produce coordinate observers, necessary for the Cauchy evolution algorithm\cite{SYO}. These hypersurface orthogonal congruences correspond to tetrads that locally add only one off-diagonal component with respect to the tetrad that diagonalizes the stress-energy tensor which is a source of important simplification. When expressed in terms of the  original tetrads (\ref{U}-\ref{W}) the stress-energy tensor is diagonal, the only change with respect to the hypersurface orthogonal tetrads (\ref{SFU}-\ref{SFW}) is just one off-diagonal component (\ref{SE03F}) of the stress-energy tensor which is an outstanding simplifying property. We can then proceed with the explicit expression for the new tetrad to apply already known algorithms \cite{JWY}$^{-}$\cite{GC} for the evolution of spacetimes where electromagnetic fields are present. The idea would be to feed already known algorithms with the new tetrads in order to evolve four-dimensional Lorentzian spacetimes where there are dynamical interactions involving gravity and electromagnetic fields. We quote from \cite{JWYCP} ``Theories with gauge freedom, such as electromagnetism and general relativity, are said to be both  ``overdetermined'' and  ``underdetermined''. They are overdetermined because there are constraints at each time that limit the freedom of the variables that are propagated, the dynamical variables. They are underdetermined because the gauge freedom means the equations of the theory cannot determine a fully unique solution. By gauge transformations, some of the variables can be changed. These changes do not alter the intrinsic physical meaning of a solution but they nevertheless can be vital in the description and recognition of the solution. That the problem of being overdetermined need be resolved at one time only (in principle), and that the gauge freedom in changing certain variables does not disturb either the feature just mentioned or the physical uniqueness of the problem are part and parcel of the well-posedness of a Cauchy problem''.

%\bibliography{your-bib-file} % place the references here.

\end{document}